\documentclass[12pt]{article}
\begin{document}

\begin{center}

{\bf Quantum Evolution Supergenerator of Superparamagnetic System
in Discrete Orientation Model}

\vskip1cm

 V.A.Buslov

\vskip1cm

{\it Department of Computational Physics, Physical Faculty,
Saint-Petersburg State University, Saint-Petersburg, Yljanovskaja
st. 1, Russia}

\vskip1cm

email: buslov@cp.phys.spbu.ru

This work is supported by RFBR grant 00-01-00480.

\end{center}

\vskip1cm

\begin{abstract}
The supergenerator of superparamagnetic system quantum evolution
is investigated in discrete orientation model (DOM). It is shown
that the generator is $J$-self-adjoint one at the case of
potential drift field agreed upon magnetic anisotropy of the
sample investigated. Perturbation theory is used for spectral
analysis. The qualitative dependence of resonance absorption
spectrum on the relation between quantum and stochastic parameters
is demonstrated.
\end{abstract}

keywords: superoperator, superparamagnetic, \"{M}ossbauer
spectroscopy

\section*{Introduction}

Qualitative and quantitative description of relaxation M\"ossbauer
spectra is based on choosing of some relaxation process model. In
the most general model of motion of magnetic moment $\vec M(t)$
suggested by Brown \cite{B} this motion is considered like
diffusion process on radius $\vert \vec M \vert $ sphere in a
drift field stipulated by the magnetic anisotropy of the pattern.
Under Born's approximation M\'ossbauer spectra line shape one can
express in terms of resolvent of the generator of quantum
evolution operator of a nucleus, averaged along all diffusion
process $\vec M(t)$ trajectories. But computation of averaged
evolution operator is connected with solving of complicated
differential equation system of partial derivatives and can not be
done in analytical form. To avoid this difficulty phenomenological
models are used, where magnetic moment $\vec M(t)$ position
becomes discrete, and process of $\vec M(t)$ motion itself is
replaced by Markovian process with final number of states, which
is described by stochastic matrix with phenomenological transition
probabilities of the magnetic moment $\vec M(t)$ from one state
(of easy magnetization) to another (so called discrete orientation
model (DOM)). In this case the problem of computation of averaged
evolution operator obtains algebraic character and reduces to
inversion of special matrices, which properties and order are
determined by the number of easy magnetization directions, the
structure of nucleus magnet electron system and by the symmetry of
pattern. In this direction certain progress was reached for SP
particles with cubic symmetry \cite{AO}. Direct account of
symmetry in frames of phenomenological DOM permitted authors
\cite{AO} to reduce the calculation of absorption line shape of SP
particles to inversion some matrices of 8- and 6- order for cubic
symmetry systems with negative and positive magnetic anisotropy
constants respectively.

For the case of large values of reduced barrier $\alpha=KV/kT$ ,
where $K$ is magnetic anisotropy constant, $k$ - Boltzmann
constant, $V$ - volume of SP particle, $T$ - temperature, DOM
parameters one can calculate by the lowest eigenvalues of
Fokker-Plank equation \cite{BMT}, \cite{BMM}, giving evolution of
$\vec M(t)$ in diffusion model \cite{BMP}, and line shape
calculated in such a way is the leading term under $\alpha
\rightarrow\infty$ of line shape calculated on the base of
diffusion model \cite{M}. For accurate calculation of the line
shape error under substitution of diffusion model to discrete one
(DOM) first of all detailed analysis of the generator of averaged
evolution operator of a nucleus in frames of DOM is necessary.

In present paper, developing symmetry considerations stated at
\cite{AO}, we investigate the structure of complete averaged
evolution operator corresponding to DOM for cubic symmetry (with 6
and 8 easy magnetization directions on $Fe^{57}$ nucleus) and some
others, maintain it's spectral analysis in frames of perturbation
theory under assumption of slow relaxation and also describe
qualitative picture of spectrum behavior in general case (subject
to temperature). In proper basis the generator matrix is composed
of equal blocks of $8\times 8$ older differing only by its
diagonal elements composed by combinations of eigenvalues of the
stochastic matrix giving DOM. Nondiagonal part of such block is
determined only by the nuclear system structure and does not
depend neither on stochastic variables nor on crystal lattice
symmetry type (on magnetic anisotropy constant sign for cubic
symmetry).

\section{Basic Formulas}

In DOM magnetic moment evolution is defined by matrix
$P=S^\texttt{T}-I$, where $S$ is stochastic matrix of transition
probabilities between easy magnetization states in unit of time.
Vector $\vec g(t)$ of magnetic moment $\vec M(t)$ distribution is
governed by ordinary differential equation

\begin{equation} {d\vec g \over dt}=P\vec g, \label{1} \ \ \ \ \vec g =(g_1,g_2,\cdots
,g_N)^\dagger, \end{equation}   where $N$ - the number of process
states (easy magnetization directions). In the case of cubic
symmetry $N$ is equal either 6 or 8 for positive or negative
magnetic anisotropy constant respectively. Matrix $P$ is naturally
to call as Markovian one.

Line shape is determined by expression \cite{AO},\cite{BP}

\begin{equation} \varphi (\omega )= {2\over \Gamma } Re \int\limits^{\infty
}_{0} e^{(i\omega - \Gamma /2)t} Sp ({\bf V}^{*}\overline
{G(t)}{\bf V}\rho ) dt , \label{2} \end{equation}   where $\omega
$ is the electromagnetic frequency, $\Gamma$ - natural half-width,
$\rho$ - the initial state density matrix, ${\bf V}$ - the
operator of interaction of the nucleus with electromagnetic field,
$\overline{G(t)}$ - the result of averaging of the quantum
evolution superoperator of the nucleus $exp(i\int^t_0 \hat
Ld\tau)$ along all diffusion process trajectories, in our case
along trajectories of Markovian process (\ref{1}) with finite
number of states. In this connection $\overline {G(t)}$ is the
solution of equation \cite{BP}:

\begin{equation} {d\overline G \over dt}=\hat P\overline G-i\hat L\overline G.
\label{3} \end{equation}   Here $\hat L$ is diagonal on stochastic
variables Liouville's superoperator, which action on the spin
transition operators $B$ is determined by the rule: $\hat
LB=H^eB-BH^g$. Here $H^{e,g}=A^{e,g}\times (\vec I^{e,g},\vec m),\
\vec I$  spin operator of a particle (for $Fe^{57}: I^g={1\over
2}, I^e={3\over 2}),\ \vec m$ - unit vector of particle
magnetization direction, $A^{e,g}$ - hyperfine interaction
constants, $\hat P=P\otimes E_q$ - operator matrix diagonal on
nuclear variables, $E_q$ - identity superoperator acting in
quantum states space.

Calculating integral (\ref{2}) taking into account (\ref{3}) one
can get the following expression for line shape \cite{AO}

\begin{equation} \varphi (\omega )=Im\sum_{a,b}^N \rho_a\langle \vec
\eta^*J(a\mid \hat L+i\hat P-\lambda \hat E\mid ^{-1}b)\vec \eta
J^\dagger \rangle , \label{4} \end{equation}   where $\lambda
=\omega +i {\Gamma \over 2} \ ,\ \Gamma$ - natural half-width,
$\vec \eta$ - vector of polarization of falling $\gamma
$-radiation, $J$ - nuclear current operator, responsible for
transitions between sublevels of ground $\vert m_g\rangle $ and
excited $\vert m_e\rangle $ nucleus states, $\rho _a $ - relative
probabilities of population of electron states $a$ (there are $N$
such states). So such  calculation is reduced to spectral analysis
of the generator of quantum evolution superoperator of a nucleus
$\hat G^{-1}(\lambda )= \hat L+i\hat P-\lambda \hat E \ $, acting
in space of operator-functions of the form: $\vert \Psi \rangle
=\vert m_e\rangle \langle m_ g\parallel a)$, which is a tensor
product of the spin operators space and stochastic space of
magnetic moment directions.  Superoperator $\hat L$, responsible
for hyperfine decomposition of levels, acts by the rule

\begin{equation} \langle \Psi\mid \hat L\mid \Psi '\rangle =[H^e_{m_em_e'}
\delta _{m_gm_g'}- H^g_{m'_gm_g} \delta _{m_em_e'} ] \delta
_{aa'}. \label{5} \end{equation}   Operator matrix $\hat P$ is
diagonal on nuclei variables

\begin{equation} \langle \Psi \mid \hat P \mid \Psi '\rangle =p_{aa'}
\delta_{m_em_e'} \delta_{m'_gm_g} \ , \label{6} \end{equation}

$$p_{aa}=-\sum_{a' \not= a}(a\mid \hat P \mid a') \ .
$$
Relaxation matrix elements $p_{aa'}$ are transition probabilities
in unit of time from state $a$ to state $a'$.

Of course one can consider  the state $| \Psi >$ at (\ref{5}) and
(\ref{6}) as usual state in direct product space of variables $
m_e$, $m_g $, $a$ and forget the superoperator nature of $\hat G
 =(\hat L +i\hat P - \lambda \hat E )^{-1}$.

\section{Spectral Analysis of Stochastic Matrices}

Let us suppose that easy magnetization axes coincide with axes of
3(4)-order of cube for the case of negative (positive) magnetic
anisotropy constant. We  accept the following indexing of these
axes (fig.1,2). For such indexing stochastic matrices of the
Markovian process, giving magnetic moment motion, look in a
following way
$$
P^{(6)}=\bordermatrix{&1&2&3&\bar 1&\bar 2&\bar 3\cr
                1&-4p-q&p&p&q&p&p\cr
                2&p&-4p-q&p&p&q&p\cr
                3&p&p&-4p-q&p&p&q\cr
                \bar 1&q&p&p&-4p-q&p&p\cr
                \bar 2&p&q&p&p&-4p-q&p\cr
                \bar 3&p&p&q&p&p&-4p-q\cr} ,
$$
$$
P^{(8)}=\bordermatrix{&1&2&3&4&\bar 1&\bar 2&\bar 3&\bar 4\cr
                  1&s&p&q&p&r&q&p&q\cr
                  2&p&s&p&q&q&r&q&r\cr
                  3&q&p&s&p&p&q&r&q\cr
                  4&p&q&p&s&q&p&q&r\cr
                  \bar 1&r&q&p&q&s&p&q&p\cr
                  \bar 2&q&r&q&p&p&s&p&q\cr
                  \bar 3&p&q&r&q&q&p&s&p\cr
                  \bar 4&q&p&q&r&p&q&p&s\cr} ,
$$
$$s=-3p-3q-r.$$
Spectral analysis of such matrices is well known
\cite{AO},\cite{BMPS},\cite{BPSK}. In particular $P^{(8)}$ has two
single eigenvalues and two triple ones ($\rho =\lambda_1=0$, $\eta
=\lambda_2=\lambda_3=\lambda_4=-2(p+2q+r)$, $\xi =
\lambda_5=\lambda_6=\lambda_7=-4(p+q)$, $\zeta
=\lambda_8=-(6p+4q)$), and $P^{(6)}$ has one single, one double
and one triple eigenvalues ($\rho = \lambda_1=0$, $\zeta
=\lambda_2=\lambda_3=-6p$, $\zeta =\lambda_4=
\lambda_5=\lambda_6=-4p-2q$). In connection with degeneration
there is some arbitrariness in choice of eigenvalues, which can be
eliminated by the requirement of definite symmetry of eigenvectors
with respect to turns around quantization axis $z$ (quantization
axis direction for 3-axial case is different from \cite{AO} and
this is connected with natural for giving case symmetry). In the
case of $N=8(6)$ in the capacity of such eigenvectors let us
choose states, that under turning at angle ${ \pi \over 2}({2\pi
\over 3})$ around quantization axis $z$ obtain factor $c_n=i^n,
n=0,1,2,3$($=exp({2\pi in\over 3}), n=0,1,2)$. In the case of
$N=8$ such vectors are the following
$$
K_n={1\over 4}[\vert 1)+c_n\vert 2)+c_n^2\vert 3)+ c_n^3\vert 4)+
\vert \bar 1)+c_n\vert \bar 2)+c_n^2\vert \bar 3)+ c_n^3\vert \bar 4)],
$$
$$
\tilde K_n={1\over 4}[\vert 1)+c_n\vert 2)+c_n^2\vert 3)+ c_n^3\vert 4)-
\vert \bar 1)-c_n\vert \bar 2)-c_n^2\vert \bar 3)- c_n^3\vert \bar 4)],
$$
and in the case $N=6$
$$
K_n={1\over \sqrt{12}}[\vert 1)+c_n\vert 2)+c_n^2\vert 3)+
\vert \bar 1)+c_n\vert \bar 2)+c_n^2\vert \bar 3)],
$$
$$
\tilde K_n={1\over \sqrt{12}}[\vert 1)+c_n\vert 2)+c_n^2\vert 3)-
\vert \bar 1)-c_n\vert \bar 2)-c_n^2\vert \bar 3)].
$$
In this connection matrix $P^{(8)}$ is diagonal in basis
$(K_0,K_1,K_2,K_3,\tilde K_0, \tilde K_1, \tilde K_2, \tilde K_3)$ and
takes form
$$
P^{(8)}={1\over 2}diag(\rho ,\xi ,\xi ,\xi ,\eta ,\eta ,
\zeta ,\eta ),
$$
and matrix
$$P^{(6)}={1\over 2}diag(\rho ,\zeta ,\zeta ,
\xi ,\xi ,\zeta )$$
 in basic $(K_0,K_1,K_2,\tilde K_0, \tilde K_1, \tilde K_2,)$.

It should be noted, that strictly speaking the use of probability
transition matrices in produced form is certain assumption, i.e.,
is adjustment to real continuous diffusion process. However, in
the case of small diffusion (or, that is the same, diffusion at
strong fields) these matrices one can get directly from continuous
diffusion model \cite{BMT},\cite{BMM} on the base of strict
asymptotic analysis of the diffusion operator. In addition it is
found that nonzero transition probabilities are only those, that
correspond to transition between potential wells, having
attraction domains with common boundary. For matrix $P^{(8)}$ this
means that $q=r=0$, and for $P^{(6)}$ that $q=0$. Moreover,
calculations show that one can obtain symmetrical Markovian
matrices only for the case of potential drift field.
Non-symmetrical field automatically means that transition
probability from one easy magnetization direction to another is
not equal to the inverse one.

\section{Reduction of Superoperator $\hat G^{-1}$ to Block Form}

Since superoperator $\hat G$ is invariant with respect to group
$C_4(C_3)$ action, it is necessary to study transformations
properties of complete space of quantum-stochastic states
respectively on such action, i.e., to construct basic elements of
complete space of states, that under turning of coordinate system
on angle ${\pi \over 2} ({2\pi \over 3})$ obtain factor $c_n$.
According to selection rules \cite{LL} superoperator $\hat G$ in
such basis is of block structure. For this aim it is naturally in
quantum variables space to extract states with such properties
relatively group $C_4(C_3)$ (such kind of states in stochastic
variables space we have constructed above) and than to construct
4(3) nonoverlapping subspaces of complete quantum-stochastic
space, the states of which obtain factor $c_n$ under turning at
angle ${\pi \over 2}({2\pi \over 3})$ around quantization axis
$z$. How it is shown in \cite{AO}, invariant with respect to group
$C_4$ subspace in the case $N=8$ is divided into two
nonoverlapping subspaces with different properties relatively time
inversion (symmetry and asymmetry subspaces). Really, all others
subspaces also suppose such division both in 4- and 3-axial cases
and also at other symmetries under proper disposition (consistent
with the symmetry) of coordinate system. So it is possible to
construct basis in which superoperator $\hat G$ consists of 8(6)
equal blocks of order $8\times 8$, that differ from each other
only by diagonal elements, composed of stochastic matrix
eigenvalues combinations. Nondiagonal part of each block is
determined only by the transition structure of nuclear system and
does not depend on sign of magnetic anisotropy constant. Denoting
by $\{ n\pm \} $ blocks ${symmetrical} \choose{asymmetric}$ with
respect to time reversal, that obtain factor $c_n=i^n, \
n=0,1,2,3\ (exp({2\pi in \over 3}),\  n=0,1,2)$ for 4(3)-axial
case and denoting quantum transition operators
$$
\alpha =\vert {1\over 2}\rangle \langle {1\over 2} \vert , \
\beta =\vert {-1\over 2}\rangle \langle {-1\over 2} \vert ,  \
\gamma = \vert {3\over 2}\rangle \langle {1\over 2} \vert ,  \
\sigma =\vert {-3\over 2}\rangle \langle {-1\over 2} \vert ,
$$
$$
\delta =\vert {1\over 2}\rangle \langle {-1\over 2} \vert , \
\varepsilon =\vert {-1\over 2}\rangle \langle {1\over 2} \vert , \
\mu =\vert {3\over 2}\rangle \langle {-1\over 2} \vert , \
\nu =\vert {-3\over 2}\rangle \langle {1\over 2} \vert ,
$$
we have the following basic functions for blocks $\{ n\pm \} $
in 4-axial case
$$
\vert 1\rangle =K_n(\alpha \mp \beta) ,\
\vert 2\rangle =K_{n+3} \delta \pm K_{n+1} \varepsilon, \
\vert 3\rangle =K_{n+3} \gamma \pm K_{n+1} \sigma , \
$$
$$
\vert 4\rangle =K_{n+2} (\mu \mp \nu), \
\vert 5\rangle =\bar K_{n+2} (\mu \pm \nu), \
\vert 6\rangle =\bar K_{n+3} \gamma \mp \bar K_{n+1} \sigma , \
$$
$$
\vert 7\rangle =\bar K_{n+3} \delta \mp \bar K_{n+1} \varepsilon, \
\vert 8\rangle =\bar K_n( \alpha \pm \beta ),
$$
and in 3-axial case
$$
\vert 1\rangle =K_n(\alpha \mp \beta) ,\
\vert 2\rangle =K_{n+2} \delta \pm K_{n+1} \varepsilon, \
\vert 3\rangle =K_{n+2} \gamma \pm K_{n+1} \sigma , \
$$
$$
\vert 4\rangle =K_{n+1} \mu \mp K_{n+2} \nu , \
\vert 5\rangle =\bar K_{n+1} \mu \pm K_{n+2} \nu, \
\vert 6\rangle =\bar K_{n+2} \gamma \mp \bar K_{n+1} \sigma , \
$$
$$
\vert 7\rangle =\bar K_{n+2} \delta \mp \bar K_{n+1} \varepsilon, \
\vert 8\rangle =\bar K_n( \alpha \pm \beta ),
$$
Here in the 4(3)-axial cases all indexes are taken by $mod4(3)$.
In this basis diagonal submatrix $P_{n\pm }$ takes form  for
4-axial case

\begin{equation}
P_{0\pm }=-diag(\Lambda ,\Omega ,\Omega ,\Omega ,\Xi ,\Upsilon ,
\Upsilon ,\Upsilon ) \label{7} \end{equation}
$$
P_{1,3\pm }=-diag(\Omega ,{\Omega \over 2}, {\Omega \over 2},
{\Omega \over 2},\Upsilon ,\Theta ,\Theta ,\Upsilon )
$$
$$
P_{2\pm }=-diag(\Omega ,\Omega ,\Omega ,\Lambda ,\Upsilon ,\Upsilon ,
\Upsilon ,\Xi ),
$$
where $\Lambda =0,\ $ $\Omega =4p+4q,\ $ $\Xi =6p+2r,\ $
$\Upsilon =2p+4q+2r,\ $ $\Theta =4p+2q+2r$. For $N=6$ we have

\begin{equation}
P_{0\pm }=-diag(\Lambda ,\Omega ,\Omega ,\Omega ,\Upsilon
,\Upsilon , \Upsilon ,\Upsilon), \label{8} \end{equation}
$$
P_{1,2\pm }=-diag(\Omega ,{\Omega \over 2}, {\Omega \over 2},
{\Omega \over 2},\Upsilon ,\Upsilon ,\Upsilon ,\Upsilon ),
$$
where $\Lambda =0,\ $ $\Omega =6p,\ $ $\Upsilon =4p+2q.\ $

Superoperator $\hat L$ action at all received subspaces
independently on symmetry appears to be the same and is given by
matrix
$$
\hat L=\left (\matrix{0&D\cr
              D^{\dagger }&0\cr} \right),
$$
$$
  D=\left(\matrix{
0&A^e/\sqrt 2&(2A^e-A^g)/\sqrt 6&(A^e-A^g)/ 2\sqrt 3\cr
A^e/\sqrt 2&0&(A^e+A^g)/2\sqrt 3&-(A^g+2A^e)/\sqrt 6\cr
-A^g/\sqrt 6&(3A^e-A^g)/2\sqrt 3&0&A^e/\sqrt 2\cr
(3A^e+A^g)/2\sqrt 3&-A^g/\sqrt 6&A^e/\sqrt 2&0\cr}\right).
$$
Block of form $D$ was introduced first in \cite{AO} for invariant
subspace of quantum-stochastic states antisymmetrical with respect
to time inversion (we denote this subspace as {0-}) in the case
$N=8$. Remarkably, that other blocks of superoperator $\hat L$
(that have another symmetry, i.e., that obtain factor $c_n$ under
turning of coordinate system on angle ${\pi \over 2}({2\pi \over
3}$)) have the same form. Moreover, one can show that the same
form of elementary block remains for certain other symmetries, for
example for tetrahedron symmetry, cylindrical symmetry and some
others. So elementary block $D$ is in some sense universal.

\section{Spectral Analysis of Quantum-Stochastic Superoperator}

Last paragraph is devoted to reduction of quantum-stochastic
superoperator to block-diagonal form convenient for calculations
its spectral characteristics (eigenvalues and eigenvectors). This
problem can be solved under assumption that one of terms $\hat L$
or $\hat P$ composite superoperator $\hat G^{-1}$ is small in
comparison with another. It is naturally to assume smallness of
stochastic term $\hat P$. Note that eigenvalues of unperturbed
superoperator $\hat L$ are known explicitly. These eigenvalues are
the super-fine  transition frequencies $\lambda_{\pm i}=\pm b_i$,
where $b_1={1\over 2}(3A^e-A^g)\ $ , $b_2={1\over 2}(A^e-A^g)\ $,
$b_3=-{1\over 2}(A^e+A^g)\ $, $b_4=-{1\over 2}(3A^e+A^g)\ $.
Corresponding  eigenfunctions have the form \cite{AO}
$$
\Psi_{\pm 1}={1\over 2}\vert 1\rangle +0+{1\over \sqrt 6}\vert 3\rangle  +
{1\over 2\sqrt 3}\vert 4\rangle \pm ({1\over 6}\vert 5\rangle +
{\sqrt 2\over 3}\vert 6\rangle +
{1\over \sqrt 6}\vert 7\rangle +{1\over 2\sqrt 3}\vert 8\rangle )
$$
$$
\Psi_{\pm 2}={1\over \sqrt 6}\vert 1\rangle -{1\over 2\sqrt 3}\vert 2\rangle -
{1\over \sqrt 3}\vert 3\rangle +0\pm (-{1\over \sqrt 6}\vert 5\rangle -
{1\over 2\sqrt 3}\vert 6\rangle +{1\over 2}\vert 7\rangle +0)
$$
$$
\Psi_{\pm 3}={1\over 2\sqrt 3}\vert 1\rangle +{1\over \sqrt 6}\vert 2\rangle +0-
{1\over 2}\vert 4\rangle \pm ({1\over 2\sqrt 5}\vert 5\rangle -
{1\over \sqrt 6}\vert 6\rangle
+0+{1\over 2}\vert 8\rangle )
$$
$$
\Psi_{\pm 4}=0-{1\over 2}\vert 2\rangle +{1\over 2\sqrt 3}\vert 3\rangle -
{1\over \sqrt 6}\vert 4\rangle \pm ({\sqrt 2\over 3}\vert 5\rangle
-{1\over 6}\vert 6\rangle +
{1\over 2\sqrt 3}\vert 7\rangle -{1\over \sqrt 6}\vert 8\rangle ).
$$
Standard perturbation theory in the case of absence of degeneration
let us calculate eigenvalues corrections
$$
\Delta \lambda_k=\langle \Psi_k\mid i\hat P\mid \Psi_k\rangle
$$
and also eigenvectors corrections. In our case for eigenvalues
this leads to results reduced in Tab.1 of eigenvalues corrections.

{\bf Tab.1a}. {\sl Eigenvalues corrections for 4-axis case.}

\begin{tabular}{|c|c|c|c|}  \hline
&block {$0\pm $}&block {$1,3\pm $}&block {$2\pm $}\\ \hline
$\delta \lambda_1$&$i(19/9p+26/9q+r)$&$
i(31/9p+26/9q+r)$&$i(3p+10/3q+r)$\\ \hline
$\delta \lambda_2$&$i(3p+8/3q+r)$&$i(3p+8/3q+r)$&$
i(3p+4q+r)$\\ \hline
$\delta \lambda_3$&$i(3p+10/3q+r)$&$i(3p+10/3q+r)$&$
i(3p+2q+r)$\\ \hline
$\delta \lambda_4$&$i(35/9p+28/9q+r)$&$
i(23/9p+28/9q+r)$&$i(3p+8/3q+r)$\\ \hline
\end{tabular}

{\bf Tab.1b}. {\sl Eigenvalues corrections for 3-axis case.}

\begin{tabular}{|c|c|c|}  \hline
&block {$0\pm $}&block {$1,2\pm $}\\ \hline
$\delta \lambda_1$&$i(3,5p+q)$&$i(4,25p+q)$\\ \hline
$\delta \lambda_2$&$i(4p+q)$&$i(4p+q)$\\ \hline
$\delta \lambda_3$&$i(4,5p+q)$&$i(3,75p+q)$\\ \hline
$\delta \lambda_4$&$i(5p+q)$&$i(3,5p+q)$\\ \hline
\end{tabular}

\section{$J$-Selfadjointness and Qualitative Spectrum Behavior}

Reductions of operator matrix $\hat L+i\hat P$ to block-diagonal
form at mentioned basis permit to receive effective way of
calculating of line shape reduce to inversion of matrices of
$8\times 8$-order corresponding to each block. In the case of
equilibrium initial distribution resonance absorption spectrum is
described by the only block of the superoperator $\hat L+i\hat P$,
namely by that, which is invariant with respect to turns on angle
${\pi \over 2} ({2\pi \over 3})$ and antisymmetrical relatively
time inversion. But if the initial distribution is arbitrary one
it is necessary to know all blocks of the superoperator $\hat
L+i\hat P$ for line shape determination.

 The superoperator $\hat L+i\hat P$ spectrum is, generally speaking,
complex one, nevertheless, it is possible to get qualitative
information on spectrum behavior at different relations between
parameters of magnetic and electron spin systems. The superoperator
$\hat L+i\hat P$ spectrum lies at lower half-plane and is placed
symmetrically relatively imaginary axis. This follows from
$J$-selfadjointness of $\hat L+i\hat P$. Really
$$
J\Biggl(\left(\matrix{0&D\cr
                      D^{\dagger }&0\cr}\right)+i\hat P\Biggr)J=
\Biggl(\left(\matrix{0&D\cr
                      D^{\dagger }&0\cr}\right)+i\hat P\Biggr)^{\dagger },
$$
when $J$={\rm i}$diag(1,1,1,1,-1,-1,-1,-1)$.

Let us consider for illustration the case of equilibrium initial
distribution, which is described by block {0-}. At the limit case
of fast relaxation (at high temperatures) eigenvalues of this
block (just as of any other block) belong to imaginary axis, and
this case corresponds to resonance absorption spectrum consisting
of the only line of Lorentz width on the frequency $\omega $. At
diminishing of relaxation eigenvalues as before some time belong
to imaginary axis moving on it, because at fast relaxation limit
there are two single and two triple eigenvalues both in 4- and
3-axial case (\ref{7}),(\ref{8}), but due to the
$J$-selfadjointness eigenvalues can leave imaginary axis only by
pairs. At certain moment one of pairs will diverge (and also this
can be not one pair from initially existing pairs in triple
eigenvalues but one of anew formed in a process of random walk
along imaginary axis). In this connection resonance absorption
spectrum will have a form of two lines of certain width,
symmetrical relatively frequency $\omega $. Under further
diminishing of relaxation parameters new pairs of lines appear,
and the width is decreased. At limit all eigenvalues of the matrix
come down from imaginary axis and are dropped on real axis, that
on resonance absorption spectrum correspond to line shape with
peaks on frequencies $\omega +b_{\pm i}$ (frequencies $b_{\pm 4}$
are prohibited and don't appear on spectrum), the width is
decreased as far as eigenvalues approach real axis, i.e., under
diminishing of relaxation or, that is the same, of temperature.

It is necessary to note that symmetrical line shape corresponds to
potential drift field (in DOM --- to symmetrical matrices $P$).
Calculations using non-symmetrical transition probabilities
demonstrate non-symmetrical line shape with respect to main
frequency $\omega$.

Author is grateful to K.A.Makarov and B.S.Pavlov for useful
discussions. This work is supported by RFBR grant 00-01-00480.

\newpage
\begin{center}
\textbf{fig. 1. Mutual disposition of the easy magnetization
directions and coordinate system (4-axis case, N=8).}
     \end{center} {\it Values of unit vector $\vec m$ of particle
magnetization}

$ \vec m=\pm \vec m_i, \  i=1,2,3,4.$

$\vec m_1=(\sqrt {2/3},0,1/\sqrt 3),\ \vec m_2=(0,\sqrt
{2/3},1/\sqrt 3),$

$\vec m_3=(-\sqrt {2/3},0,1/\sqrt 3),\ \vec m_4=(0,-\sqrt
{2/3},1/\sqrt 3).$

\textit{Easy magnetization directions coincide with ones from the
center of cube to its vertices.}

\vskip2cm \unitlength=1.00mm \special{em:linewidth 0.4pt}
\linethickness{0.4pt}
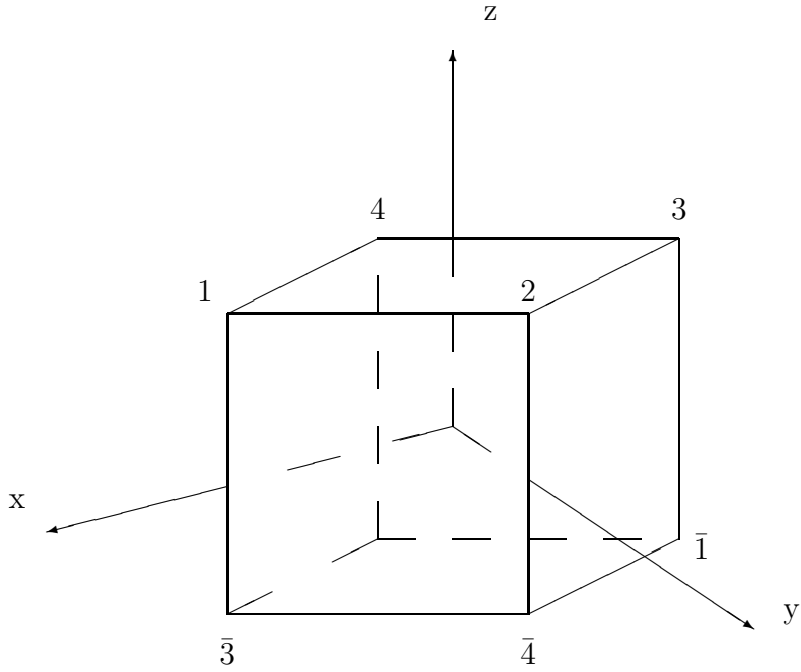
\begin{picture}(125.00,100.00)(20.00,0.00)
\put(50.00,20.00){\line(1,0){40.00}}
\put(90.00,20.00){\line(0,1){40.00}}
\put(90.00,60.00){\line(-1,0){40.00}}
\put(50.00,60.00){\line(0,-1){40.00}}
\put(90.00,20.00){\line(2,1){20.00}}
\put(110.00,30.00){\line(0,1){40.00}}
\put(110.00,70.00){\line(-2,-1){20.00}}
\put(50.00,60.00){\line(2,1){20.00}}
\put(70.00,70.00){\line(1,0){40.00}}
\put(70.00,65.00){\line(0,-1){5.00}}
\put(70.00,55.00){\line(0,-1){5.00}}
\put(70.00,45.00){\line(0,-1){5.00}}
\put(70.00,35.00){\line(0,-1){5.00}}
\put(70.00,30.00){\line(1,0){5.00}}
\put(80.00,30.00){\line(1,0){5.00}}
\put(90.00,30.00){\line(1,0){5.00}}
\put(100.00,30.00){\line(1,0){5.00}}
\put(70.00,30.00){\line(-2,-1){6.00}}
\put(50.00,20.00){\line(2,1){6.00}}
\put(80.00,45.00){\line(-4,-1){8.00}}
\put(65.00,41.00){\line(-4,-1){7.00}}
\put(50.00,37.00){\vector(-4,-1){24.00}}
\put(80.00,45.00){\line(3,-2){5.00}}
\put(90.00,38.00){\vector(3,-2){30.00}}
\put(80.00,45.00){\line(0,1){5.00}}
\put(80.00,55.00){\line(0,1){5.00}}
\put(80.00,65.00){\vector(0,1){30.00}}
\put(80.00,95.00){\line(0,0){0.00}}
\put(47.00,63.00){\makebox(0,0)[cc]{1}}
\put(90.00,63.00){\makebox(0,0)[cc]{2}}
\put(110.00,74.00){\makebox(0,0)[cc]{3}}
\put(70.00,74.00){\makebox(0,0)[cc]{4}}
\put(113.00,29.00){\makebox(0,0)[cc]{$\bar 1$}}
\put(90.00,15.00){\makebox(0,0)[cc]{$\bar 4$}}
\put(50.00,15.00){\makebox(0,0)[cc]{$\bar 3$}}
\put(22.00,35.00){\makebox(0,0)[cc]{x}}
\put(125.00,20.00){\makebox(0,0)[cc]{y}}
\put(85.00,100.00){\makebox(0,0)[cc]{z}}
\end{picture}
\newpage
\begin{center}
\textbf{fig. 2. Mutual disposition of the easy magnetization
directions and coordinate system (3-axis case, N=6).}
\end{center}

{\it Values of unit vector $\vec m$ of particle magnetization}

$ \vec m=\pm \vec m_i, \  i=1,2,3.$

$\vec m_1=(\sqrt {2\over 3},0,{1\over \sqrt 3}),\ \vec
m_2=({-1\over \sqrt 6},{1\over \sqrt 2},{1\over \sqrt 3}),\ \vec
m_3=({-1\over \sqrt 6},{-1\over \sqrt 2},{1\over \sqrt 3}).$

\textit{Easy magnetization directions coincide with ones from the
center of cube to the centers of its faces. } \vskip2cm
\unitlength=1.00mm \special{em:linewidth 0.4pt}
\linethickness{0.4pt}
\begin{picture}(145.00,111.00)(20.00,0.00)
 \put(80.00,20.00){\line(2,1){40.00}}
\put(120.00,40.00){\line(-2,1){20.00}}
\put(100.00,50.00){\line(-2,-1){40.00}}
\put(60.00,30.00){\line(2,-1){20.00}}
\put(60.00,30.00){\line(-1,2){20.00}}
\put(40.00,70.00){\line(2,1){40.00}}
\put(80.00,90.00){\line(1,-2){20.00}}
\put(80.00,90.00){\line(2,-1){20.00}}
\put(100.00,80.00){\line(1,-2){20.00}}
\put(80.00,55.00){\line(0,1){5.00}}
\put(80.00,65.00){\line(0,1){5.00}}
\put(80.00,75.00){\line(0,1){5.00}}
\put(80.00,85.00){\vector(0,1){26.00}}
\put(80.00,20.00){\line(-1,2){5.00}}
\put(70.00,40.00){\line(-1,2){5.00}}
\put(60.00,60.00){\line(-2,1){10.00}}
\put(60.00,60.00){\line(2,1){10.00}}
\put(80.00,70.00){\line(2,1){10.00}}
\put(80.00,55.00){\line(3,-1){6.00}}
\put(93.00,51.00){\line(3,-1){8.00}}
\put(110.00,45.00){\vector(3,-1){30.00}}
\put(80.00,55.00){\line(-3,-2){8.00}}
\put(65.00,45.00){\vector(-3,-2){30.00}}
\put(70.00,60.00){\circle*{2.00}}
\put(100.00,65.00){\circle*{2.00}}
\put(90.00,35.00){\circle*{2.00}} \put(70.00,75.00){\circle{2.00}}
\put(60.00,45.00){\circle{2.00}} \put(90.00,50.00){\circle{2.00}}
\put(85.00,110.00){\makebox(0,0)[cc]{z}}
\put(30.00,25.00){\makebox(0,0)[cc]{x}}
\put(145.00,35.00){\makebox(0,0)[cc]{y}}
\put(74.00,56.00){\makebox(0,0)[cc]{1}}
\put(104.00,61.00){\makebox(0,0)[cc]{2}}
\put(94.00,35.00){\makebox(0,0)[cc]{$\bar 3$}}
\put(65.00,72.00){\makebox(0,0)[cc]{3}}
\put(57.00,48.00){\makebox(0,0)[cc]{$\bar 2$}}
\put(86.00,47.00){\makebox(0,0)[cc]{$\bar 1$}}
\end{picture}

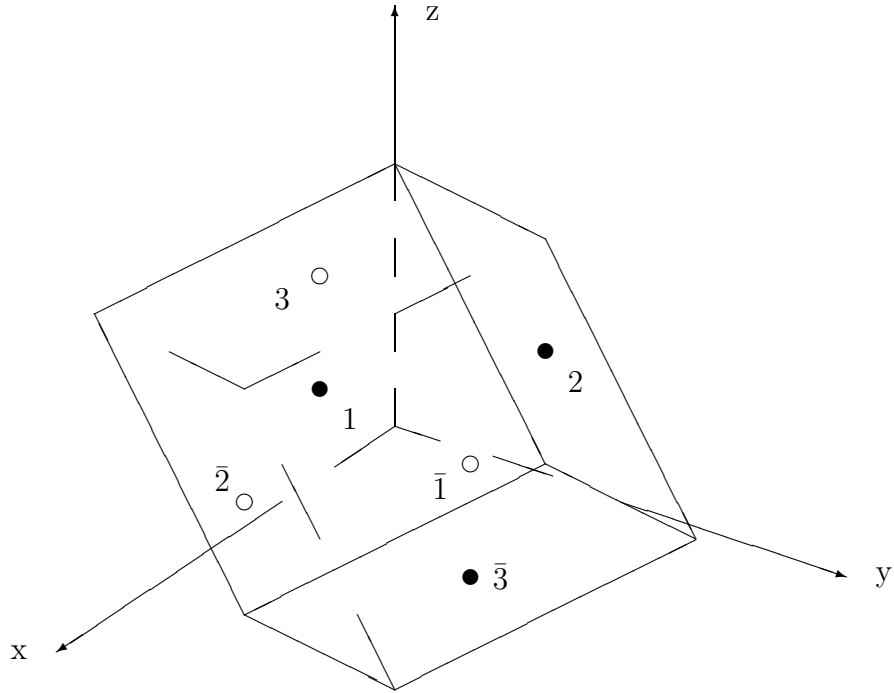
\end{document}